# Reconstructing Transportation Cost Planning Theory: A Multi-Layered Framework Integrating Stepwise Functions, AI-Driven Dynamic Pricing, and Sustainable Autonomy

**Samuel Darwisman**

*Institut Transportasi dan Logistik Trisakti, Jakarta, Indonesia*
*Email: 23c606001088@student.itltrisakti.ac.id*

**Abstract**

Context: The theoretical landscape of transportation cost planning is shifting from deterministic linear models to dynamic, data-driven optimization. As supply chains face volatility, static cost assumptions prove increasingly inadequate. Gap: Despite rapid technological advancements, a unified framework linking economic production theory with the operational realities of autonomous, sustainable logistics remains absent. Existing models fail to address non-linear stepwise costs and real-time stochastic variables introduced by market dynamics. Objective: This study reconstructs transportation cost planning theory by synthesizing Grand, Middle-Range, and Applied theories. It aims to integrate stepwise cost functions, AI-driven decision-making, and environmental externalities into a cohesive planning model. Method: A systematic theoretical synthesis was conducted using 28 high-impact papers published primarily between 2018 and 2025, employing multi-layered analysis to reconstruct cost drivers. Results: The study identifies three critical shifts: the transition from linear to stepwise fixed costs, the necessity of AI-driven dynamic pricing for revenue optimization, and the role of Autonomous Electric Vehicles (AEVs) in minimizing long-term marginal costs. Conclusion: A "Dynamic-Sustainable Cost Planning Theory" is proposed, arguing that cost efficiency now depends on algorithmic prediction and autonomous fleet utilization rather than simple distance minimization.

**Keywords:** *Transportation Cost Planning, Stepwise Costs, Dynamic Pricing, Artificial Intelligence, Green Logistics, Autonomous Electric Vehicles.*

# 1. INTRODUCTION

Transportation serves as the circulatory system of the global economy, determining the velocity, reliability, and cost-efficiency of trade distribution networks. It accounts for a significant portion of the total landed cost of goods, often ranging from 40% to 60% of logistics expenditures depending on the industry and geography. For decades, the theoretical underpinnings of transportation cost planning have been deeply rooted in classical Production Theory. Hitchcock (1941) laid the seminal mathematical groundwork by formulating the classical distribution problem, which sought to minimize the linear cost of shipping homogenous goods from multiple sources to numerous destinations. This foundational work established the "Transportation Problem" as a cornerstone of Operations Research (OR), focusing primarily on the geometric optimization of flow. Later, Crainic and Laporte (1997) refined this foundation by structuring transportation planning into a rigid hierarchy of strategic (long-term network design), tactical (service frequency and scheduling), and operational (real-time routing) levels. In this classical view, cost planning was largely a deterministic exercise where managers calculated fixed costs, such as vehicles and facilities, and variable costs, like fuel and labor, based on historical averages to project future expenditures. The primary objective was minimization, finding the lowest cost path through a static network graph under capacity constraints. This approach assumed a relatively stable environment where input prices were predictable, and regulatory frameworks were static.

However, the operational environment of the 2020s differs radically from the mid-20th century. The advent of Industry 4.0 and the burgeoning Industry 5.0 has introduced a level of data granularity and real-time variability that classical linear models are ill-equipped to handle. Wheat, Odolinski, and Smith (2022) articulate that production theory in transportation must now evolve to encompass strategic operational insights, complex regulatory structures, and the competitive pressures of deregulated markets. They argue that the "production" of transport services is no longer a simple linear function of distance and mass (ton-kilometers) but involves complex efficiency frontiers influenced by technology adoption, policy compliance, and market structure. The rise of e-commerce, the volatility of fossil fuel prices, and the urgent mandate for decarbonization have transformed transportation from a cost center into a strategic differentiator. Despite these advancements in general production theory and operations research, a significant theoretical gap persists in how transportation costs specifically are modeled and managed in the digital era. Traditional accounting and planning frameworks continue to treat fixed costs as static overheads and variable costs as linear functions of output. This oversimplification fails to capture the granular realities of modern logistics, leading to significant budget variances and operational inefficiencies.

Mostafa, Moustafa, and Elshaer (2023) provide empirical evidence that fixed costs in two-stage sustainable supply chains are, in fact, dynamic variables that significantly influence network optimization. Their study suggests that when fixed costs rise—for instance, due to capital investment in green technology or autonomous fleets—the entire optimal configuration of the supply chain shifts, challenging the static assumptions of classical theory. Furthermore, the externalization of environmental costs represents a critical failure in traditional planning. Litman (2020) argues that ignoring indirect costs and non-market externalities—such as carbon emissions, congestion, noise pollution, and accident risks—leads to suboptimal transport decisions. While "Green Logistics" has emerged as a concept, it is often treated as a constraint or a secondary consideration rather than an integral component of the cost function itself. There is a distinct lack of a unified theory that treats environmental impact, autonomous technology, and algorithmic pricing as endogenous variables in the cost planning equation. Current literature often addresses these elements in isolation—discussing AI in one silo, sustainability in another, and cost accounting in a third—without synthesizing them into a cohesive planning framework. This fragmentation prevents practitioners from fully leveraging the potential of digital transformation to optimize Total Cost of Ownership (TCO).

The primary objective of this paper is to reconstruct Transportation Cost Planning Theory by synthesizing insights from Grand, Middle-Range, and Applied theories. Specifically, this study aims to deconstruct the limitations of linear cost models in the face of modern supply chain complexities, particularly the non-linearity of capacity acquisition. Furthermore, it seeks to integrate the concept of Stepwise Cost Functions to reflect the non-linear reality of fleet capacity management, where costs jump at discrete thresholds. Additionally, the research establishes the role of Artificial Intelligence (AI) and Machine Learning (ML) not merely as operational tools, but as foundational elements of dynamic pricing and real-time cost control mechanisms. Finally, the study formulates a "Dynamic-Sustainable Cost Planning Theory" that internalizes environmental costs through the adoption of Autonomous Electric Vehicles (AEVs) and collaborative logistics networks. This study contributes to the academic discourse by bridging the disparate fields of operations research, computer science, and environmental economics. By analyzing high-impact literature published primarily between 2018 and 2025, we propose a theoretical framework that is robust enough to handle the stochastic nature of the modern digital economy. We validate the transition to stepwise costs using recent mathematical proofs (Liu, 2025), demonstrate the efficacy of AI in real-time revenue management (Ghaffari et al., 2025), and quantify the economic benefits of sustainable autonomy (Zhang et al., 2025). This paper moves the field from a retrospective, accounting-based view of costs to a prospective, algorithm-driven perspective, offering a roadmap for future research and practice in sustainable transportation planning.

## 2. THEORETICAL FRAMEWORK

To reconstruct the theory of transportation cost planning effectively, this study adopts a hierarchical theoretical approach. We analyze cost drivers through three distinct layers: Grand Theory (Macro), Middle-Range Theory (Meso), and Applied Theory (Micro). This structure allows for a comprehensive understanding of how high-level economic principles translate into specific operational algorithms and managerial decisions. At the macro level, transportation cost planning is grounded in Production Theory. Classical production theory posits that firms seek to maximize output for a given set of inputs, or minimize costs for a given level of output. In transportation, the "output" is typically defined as the movement of goods or passengers over distance (ton-kilometers or passenger-kilometers). Wheat et al. (2022) provide a modern critique and expansion of this Grand Theory within the context of the Handbook of Production Economics. They argue that in the transportation sector, production functions are complicated by network effects, economies of density, and regulatory constraints. Unlike a factory where inputs (raw materials) are converted to outputs (products) in a controlled environment, transportation production occurs in a shared, open network where external factors (congestion, weather, regulation) heavily influence the input-output ratio. Wheat et al. (2022) suggest that modern production theory must account for "allocative efficiency"—the ability to combine inputs in optimal proportions given their respective prices. In the context of this study, this implies that the "inputs" are no longer just fuel, labor, and vehicles, but also data, algorithms, and autonomy. The efficiency frontier is shifting; firms that fail to adopt digital inputs will operate below the production possibility frontier, incurring higher relative costs. This Grand Theory provides the economic justification for why firms must continuously seek cost optimization through structural transformation (e.g., adopting AI and AEVs) rather than simple operational cutting. It reframes transportation not as a cost center, but as a value-generating activity where efficiency is derived from information symmetry and asset utilization.

Descending to the meso-level, two middle-range theories are critical for reconstructing cost planning: the Theory of Stepwise Costs and Sustainable Supply Chain Management (SSCM). Classical economic models often assume variable costs are continuous and linear. This implies that the cost of transporting 101 units is incrementally higher than transporting 100 units in a smooth curve. However, Liu (2025) challenges this by introducing a bounding procedure for transportation problems with stepwise costs. The core argument is that transportation capacity is not perfectly divisible. One cannot hire 1.5 trucks; one must hire either one or two. Therefore, costs behave as a "step function"—they remain constant over a specific range of capacity utilization but jump significantly when a threshold is crossed, such as the moment a new vehicle is activated or a new route is opened. Shivani, Chauhan, and Tuli (2024) further elaborate on this by applying heuristic approaches to "step fixed-charge" problems in bulk transportation. Their work demonstrates that ignoring these steps leads to significant budget variances, as the marginal cost of the unit that breaches a capacity threshold is disproportionately high. This theory is essential for understanding the "lumpiness" of transportation investment and operational expenses. It suggests that cost planning must focus on capacity saturation points to avoid unnecessary step-jumps in expenditure. Concurrently, SSCM theory serves as the ethical and regulatory framework for modern cost planning. Carter and Rogers (2008) established the foundational definition of SSCM as the strategic integration of social, environmental, and economic goals. In the context of 2024–2026, this theory has evolved from a corporate social responsibility (CSR) initiative to a core operational mandate. Javanpour et al. (2025) advance this theory by integrating carbon emission caps directly into Integer Linear Programming (ILP) models. This effectively treats environmental impact not as an externality, but as a tangible cost constraint. Under this theoretical lens, a route that is financially cheap but environmentally expensive is considered "inefficient" because it incurs latent costs (taxes, penalties, reputational damage) that will eventually materialize. This perspective compels planners to view carbon emissions as a "shadow price" that must be minimized alongside financial costs.

At the micro or operational level, Information Processing Theory via Artificial Intelligence (AI) and Digital Transformation Theory dominate the recent discourse. These applied theories explain how the grand and middle-range objectives are achieved in practice. Waller et al. (2026) conceptualize Digital Transformation in the transportation sector as the process of "unleashing the power of data." They argue that data is a new factor of production. In traditional theory, decision-making was bounded by limited information (Bounded Rationality). However, with the advent of Big Data Analytics, as described by Mageto (2021) and Ushakov et al. (2022), organizations can process vast amounts of structured and unstructured data to reduce uncertainty. This capability allows for the precise calculation of risks and the optimization of resource allocation in near real-time. Singh (2024) reinforces this by illustrating how Deep Learning (DL) and the Internet of Things (IoT) revolutionize transportation management. Applied theory in this context suggests that cost planning is a dynamic process of "predictive analytics." Instead of reacting to costs after they occur, AI models predict fuel consumption, traffic congestion, and maintenance needs before they happen, allowing for preemptive cost mitigation. For instance, predictive maintenance models can forecast component failures, preventing costly downtime and emergency repairs, which are significant deviations in traditional cost planning.

Table 1. The Hierarchical Theoretical Framework of Modern Transportation Cost Planning

| Level of Theory | Theoretical Concept | Focus of Analysis in Transportation | Key Literature Source |
|---|---|---|---|
| **Grand Theory** | Production Theory | Macro-economic efficiency, Input-Output optimization, Allocative efficiency. | Wheat et al. (2022); Hitchcock (1941) |
| **Middle-Range** | Stepwise Cost Function | Non-linear cost behavior, Capacity thresholds, Fixed-charge problems. | Liu (2025); Shivani et al. (2024); Mostafa et al. (2023) |
| **Middle-Range** | Sustainable SCM | Internalization of environmental externalities, Green Logistics. | Javanpour et al. (2025); Carter & Rogers (2008) |
| **Applied Theory** | AI & Dynamic Pricing | Real-time revenue management, Machine Learning for demand forecasting. | Ghaffari et al. (2025); Peng et al. (2025) |
| **Applied Theory** | Autonomous Operations | CapEx vs. OpEx trade-offs, Electric Fleet Grid Integration. | Zhang et al. (2025); Zhong et al. (2023) |

## 3. METHODOLOGY

This research employs a Systematic Theoretical Synthesis approach to reconstruct the transportation cost planning model. Unlike empirical studies that rely on primary data collection from a single case study, this methodology treats high-impact academic literature as the dataset, analyzing the evolution of concepts, models, and algorithms to formulate a new theoretical framework. The literature selection process was rigorous, targeting high-impact journals and conference proceedings indexed in Scopus. The selection criteria were strictly defined to ensure novelty and relevance to the digital era, with a primary focus on literature published between 2018 and 2025, including select early-access papers for 2026 to capture future-looking trends. This temporal constraint ensures that the review captures the latest developments in Industry 4.0, Artificial Intelligence (AI), and post-pandemic supply chain resilience, while seminal works such as Hitchcock (1941) and Crainic & Laporte (1997) were included solely to establish the historical baseline. Search terms utilized during the selection process included "Transportation Cost Planning," "Stepwise Costs," "Dynamic Pricing," "Machine Learning in Logistics," "Autonomous Electric Vehicles," and "Green Supply Chain." Furthermore, the types of documents selected were limited to high-quality journal articles, such as those from *Transportation Research Part E*, *European Journal of Operational Research*, and *Sustainability*, as well as authoritative book chapters, including the *Handbook of Production Economics*, and high-impact conference proceedings. A total of 28 documents were selected for deep analysis, representing a cross-section of mathematical modeling, computer science, and economic theory, thus providing a holistic view of the problem domain.

The analysis followed a three-step coding and synthesis mechanism to systematically reconstruct the theoretical framework. First, in the deconstruction phase, each selected paper was analyzed to identify specific cost drivers and planning models proposed. For example, papers on Autonomous Electric Vehicles (AEVs) were deconstructed to identify how they alter the ratio of fixed to variable costs, while papers on Artificial Intelligence (AI) were analyzed to understand how they change the speed and accuracy of decision-making. The goal of this phase was to isolate the "variables of change" that distinguish modern logistics from traditional models. Second, the identified concepts were mapped onto the theoretical hierarchy defined in the previous section. Mathematical models regarding stepwise costs, such as those by Liu (2025), were mapped to Middle-Range Theory, while AI applications, like those discussed by Ghaffari et al. (2025), were mapped to Applied Theory. This mapping ensured that every new theoretical proposition was grounded in established literature. Finally, the reconstruction phase involved synthesizing these mapped concepts into a unified framework. Relationships were drawn between disparate elements—for instance, linking the "stepwise" nature of costs to the "dynamic pricing" capabilities of AI. This synthesis process allowed for the formulation of Darwisman's Synthesis sections presented in the Results, which serve as the building blocks for the proposed "Dynamic-Sustainable Cost Planning Theory."

## 4. RESULTS

The systematic synthesis of the literature reveals three major transformations that necessitate a reconstruction of transportation cost planning theory. This section presents the empirical evidence from the literature followed by Darwisman's Synthesis, which formalizes the novel theoretical contributions of this study. Traditional transportation planning relies heavily on the assumption that variable costs increase linearly with volume. However, the analysis of recent mathematical literature proves this assumption to be obsolete for complex, modern networks. Liu (2025) provides a mathematical proof through a new bounding procedure, demonstrating that transportation costs follow a stepwise trajectory. In reality, capacity is acquired in discrete lumps, such as a truck, a container, or a driver's shift. Costs remain flat until the capacity limit of the current resource is reached, at which point the cost "jumps" to a new level to acquire additional capacity. This finding is critical because it invalidates the use of simple linear programming (LP) for accurate budget forecasting. Instead, Mixed-Integer Linear Programming (MILP) is required to model these binary decision variables of hiring or not hiring capacity. Supporting this, Shivani, Chauhan, and Tuli (2024) applied heuristic approaches to "step fixed-charge" bulk transportation problems. Their results indicate that failing to account for these steps leads to significant inefficiencies, particularly in bulk logistics where margins are thin. Furthermore, Khalilzadeh and Banihashemi (2020) argue that under uncertainty, these fixed-cost batch problems become NP-hard, requiring advanced metaheuristic algorithms to solve. This collective evidence confirms that modern cost planning must focus on capacity utilization thresholds, optimizing the "steps" rather than just the "slope" of the cost curve.

Table 2. Evolution from Linear to Stepwise Cost Planning Models

| Dimension | Classical Linear Theory | Renewed Stepwise Theory (2018–2025) | Implications for Planning |
| --- | --- | --- | --- |
| Cost Behavior | Continuous and linear relative to distance/volume. | Discontinuous, stepwise jumps based on capacity thresholds (Liu, 2025). | Managers must optimize for capacity utilization to avoid unnecessary "steps." |
| Fixed Costs | Treated as static overheads. | Dynamic variables influencing network design (Mostafa et al., 2023). | Fixed costs act as strategic decision variables, not just sunk costs. |
| Methodology | Simple Linear Programming (LP). | Mixed-Integer Programming (MIP) & Meta-heuristics (Shivani et al., 2024). | Requirement for advanced computational tools and solvers. |
| **Risk Factor** | Deterministic. | Stochastic/Uncertain (Khalilzadeh & Banihashemi, 2020). | Planning must include buffers and scenario analysis. |

**Darwisman's Synthesis 1: The Theory of Stepwise Capacity Economics**

*Based on the evidence from Liu (2025) and Shivani et al. (2024), Samuel Darwisman proposes the "Theory of Stepwise Capacity Economics." This theory asserts that cost efficiency in modern transportation is strictly a function of "full-load optimization" relative to discrete capacity steps, rather than a function of continuous distance. In classical theory, the marginal cost curve is assumed to be smooth. However, Darwisman argues that the marginal cost of transporting an additional unit is near zero until the capacity threshold (step) is reached, at which point the marginal cost becomes infinite (requiring the acquisition of a new vehicle or container). Therefore, the primary objective of cost planning shifts from "distance minimization" to "capacity alignment." Planners must focus on managing demand to fit within existing capacity steps or collaboratively sharing capacity to smooth out these steps.*

The second major result is the integration of Artificial Intelligence (AI) into the cost recovery mechanism. Cost planning is no longer just about minimizing expense; it is increasingly about maximizing revenue through Dynamic Pricing to cover dynamic costs. Ghaffari, Afsharian, and Taleizadeh (2025) present a machine learning-driven approach to dynamic pricing in multi-channel supply chains. They argue that static "cost-plus" models fail to capture the real-time fluctuations in demand and capacity. By utilizing predictive analytics, transport operators can adjust prices in real-time. For instance, if a stepwise cost increase is triggered (e.g., a new truck is needed), the AI can dynamically raise prices for spot-market shipments to cover this marginal cost. On the operational side, minimization of distance is no longer the sole objective. Peng, Jiao, and Zhang (2025) developed a Machine Learning-Powered Dynamic Fleet Routing system that focuses on real-time fuel economy. Their model incorporates smart weight sensing and intelligent traffic reasoning. Unlike traditional static routing (which plans routes at the start of the day), this system continuously re-optimizes routes based on real-time traffic data and the changing weight of the vehicle as deliveries are made. This proves that integrating real-time data reduces fuel consumption significantly more than traditional static routing. Vizhi et al. (2025) support this by demonstrating that linear regression models can predict logistics costs with high accuracy, enabling dynamic pricing strategies that protect profit margins against market volatility. Furthermore, Luo (2024) emphasizes that algorithmic research in cross-border e-commerce logistics forecasting is essential for optimizing resource allocation in dynamic environments. The capability to forecast travel times and emissions using advanced regression models, as shown by Yedla, Naidu, and Sharma (2026), further strengthens the case for AI integration.

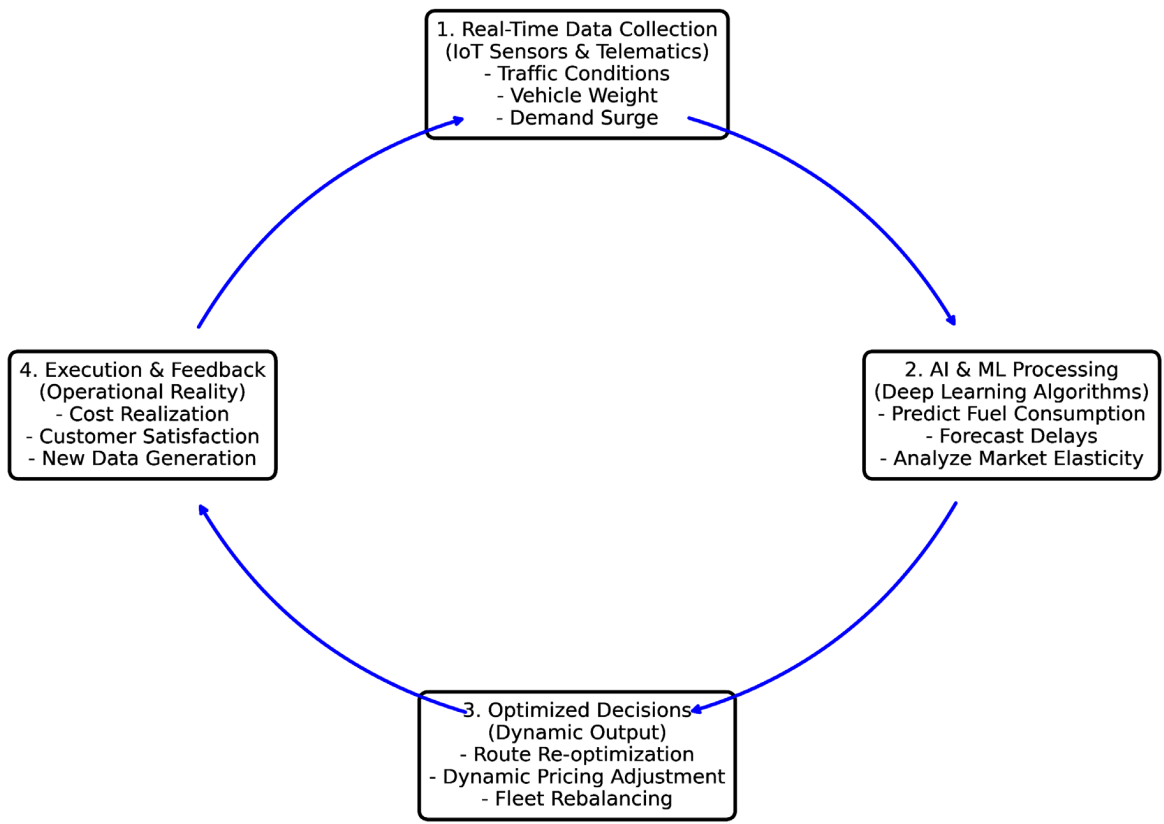

**Figure 1. The AI-Driven Dynamic Cost Planning Cycle**
*(Source: Developed by author based on Ghaffari et al., 2025; Peng et al., 2025)*

This conceptual figure illustrates the feedback loop. 1. Sensors (IoT) collect real-time data on traffic, weight, and demand. 2. AI Algorithms (Deep Learning) process this data to predict fuel consumption and delay risks. 3. The system outputs two decisions simultaneously: Optimal Route (Cost Minimization) and Dynamic Price (Revenue Maximization). 4. The Loop repeats continuously.

**Darwisman's Synthesis 2: Algorithmic Elasticity Theory**

*Synthesizing the work of Ghaffari et al. (2025) and Peng et al. (2025), Darwisman introduces the "Algorithmic Elasticity Theory." Classical economics assumes price elasticity of demand is a relatively static market property. This synthesis argues that in a digital logistics network, price elasticity is dynamic and can be manipulated through AI. By using algorithms to predict demand pulses and route conditions in real-time, planners can dynamically adjust pricing to "smooth out" demand peaks. This effectively forces volume to align with the efficient ranges of the "Stepwise*

*Capacity" described in Darwisman's Synthesis 1. Under this theory, AI serves as the "market maker," matching supply capacity with demand pulses instantly to ensure that the cost of every mile is recovered. This transforms cost planning from a defensive accounting exercise into an aggressive revenue management strategy.*

The third shift is the structural change in cost composition driven by Autonomous Electric Vehicles (AEVs). Traditional cost planning treats labor (drivers) and fuel (diesel) as the primary variable costs. However, the literature from 2023–2025 suggests a paradigm shift. Zhang et al. (2025) present a joint optimization model for fleet sizing and charging station planning for AEVs. Their findings are transformative: smart rebalancing of autonomous fleets can reduce investment costs by approximately 50%. In this model, the cost structure shifts from high variable costs (labor/fuel) to high fixed costs (technology investment/charging infrastructure). However, the marginal cost of operation drops significantly. Zhong et al. (2023) analyzed the energy and environmental impacts of shared autonomous vehicles (SAVs) under different pricing strategies. They concluded that appropriate pricing strategies for SAVs could reduce PM2.5 emissions and energy consumption by 56–64%, and up to 74% with electrification. This demonstrates that pricing policy is now a tool for emission reduction, linking environmental performance directly to economic efficiency. Moreover, Chupin et al. (2025) propose a Multi-Objective Optimization method using the NSGA-III algorithm, confirming that modern cost planning must optimize for cost, time, and emissions simultaneously.

**Table 3. Impact of Autonomous & Electric Technologies on Cost Structures**

| Cost Component | Traditional Internal Combustion Engine (ICE) | Autonomous Electric Vehicles (AEV) | Theoretical Implication |
|---|---|---|---|
| **Fuel / Energy** | High Volatility (Oil prices). | Stable (Electricity), optimized via Vehicle-to-Grid (V2G). | Shift from passive consumption to active Grid Management (Zhang et al., 2025). |
| **Labor Cost** | High Variable Cost (Drivers). | Minimized / Zero (Autonomous Systems). | Shift from Variable Operating Cost to Capital Investment cost. |
| **Maintenance** | Reactive / High cost of downtime. | Predictive / Low cost via AI and Telematics. | Integration of Reliability Theory (Ujlacka & Konečny, 2025). |
| **Emissions** | Externality (often ignored). | Internalized Cost (Carbon Caps). | Sustainability becomes an optimization objective (Zhong et al., 2023). |

**Darwisman's Synthesis 3: The Sustainable Autonomy Paradox**

*Drawing from Zhang et al. (2025) and Zhong et al. (2023), Samuel Darwisman formulates the "Sustainable Autonomy Paradox." Traditional financial theory suggests that high fixed costs (heavy capital investment) increase risk and reduce flexibility. However, Darwisman argues the opposite for Autonomous Electric Vehicles. The Sustainable Autonomy Paradox posits that the high CapEx of autonomous fleets is the very mechanism that drives long-term marginal costs (OpEx) toward zero. By eliminating labor costs (the largest variable cost component) and stabilizing energy costs via grid integration (V2G), AEVs decouple transportation costs from the volatility of human and fossil fuel markets. Therefore, to achieve true cost sustainability, organizations must embrace this paradox: they must increase their fixed asset investment to achieve near-zero marginal operational costs.*

## 5. DISCUSSION: THE DARWISMAN FRAMEWORK

The findings and syntheses presented in Section 4 coalesce into a unified theoretical framework. This section discusses the comprehensive "Dynamic-Sustainable Cost Planning Theory" proposed by Darwisman and its implications for the industry.

### 5.1. Unifying the Syntheses

The three theoretical constructs formulated by Darwisman—Stepwise Capacity Economics, Algorithmic Elasticity, and the Sustainable Autonomy Paradox—are not isolated concepts but rather form an interdependent and cohesive system. The Theory of Stepwise Capacity Economics defines the fundamental structure of modern transportation costs, characterizing them as non-linear and "lumpy" due to discrete capacity thresholds. To manage this structural complexity, the Algorithmic Elasticity Theory provides the necessary operational mechanism, utilizing Artificial Intelligence to dynamically match demand fluctuations with specific capacity steps, thereby smoothing out inefficiencies. Finally, the Sustainable Autonomy Paradox offers the long-term strategic solution to lower the overall baseline of these costs by advocating for significant capital investment in Autonomous Electric Vehicles to drastically reduce marginal operational costs. Collectively, this unified theory posits that transportation cost efficiency is no longer merely about finding the shortest path, or Distance Minimization, but rather about achieving "Synchronization"—specifically, the synchronization of fleet capacity with demand pulses using Artificial Intelligence, and the synchronization of energy consumption with grid availability through the integration of Autonomous Electric Vehicles.

**5.2. Digital Transformation as the Core Enabler**

This new theory relies heavily on digital maturity. Waller et al. (2026) assert that "unleashing the power of data" is the primary driver of transformation. Without the granular visibility provided by Big Data Analytics, as described by Mageto (2021) and Ushakov et al. (2022), the stepwise adjustments required by the theory are impossible. Wang and Liu (2025) further emphasize that the Internet of Things (IoT) is the physical enabler that connects goods and vehicles. IoT sensors provide the real-time data on vehicle weight and traffic that feed the AI models described by Peng et al. (2025). Therefore, **Darwisman** concludes that IT infrastructure investment is not an overhead but a prerequisite for cost optimization.

**5.3. Collaborative Logistics and Blockchain**

Cost reduction is maximized not in isolation, but through collaboration. Pala and Stecca (2025) introduce the concept of "Fleet Coalitions," where collaborative planning balances economic and environmental costs. By sharing transport resources, companies can reduce empty miles (a major source of waste) and lower the "stepwise" costs of adding new capacity. However, collaboration requires trust. Dutta et al. (2020) highlight Blockchain technology as the mechanism to ensure trust, security, and transparency. Blockchain enables smart contracts that automatically allocate costs between coalition partners, reducing administrative friction.

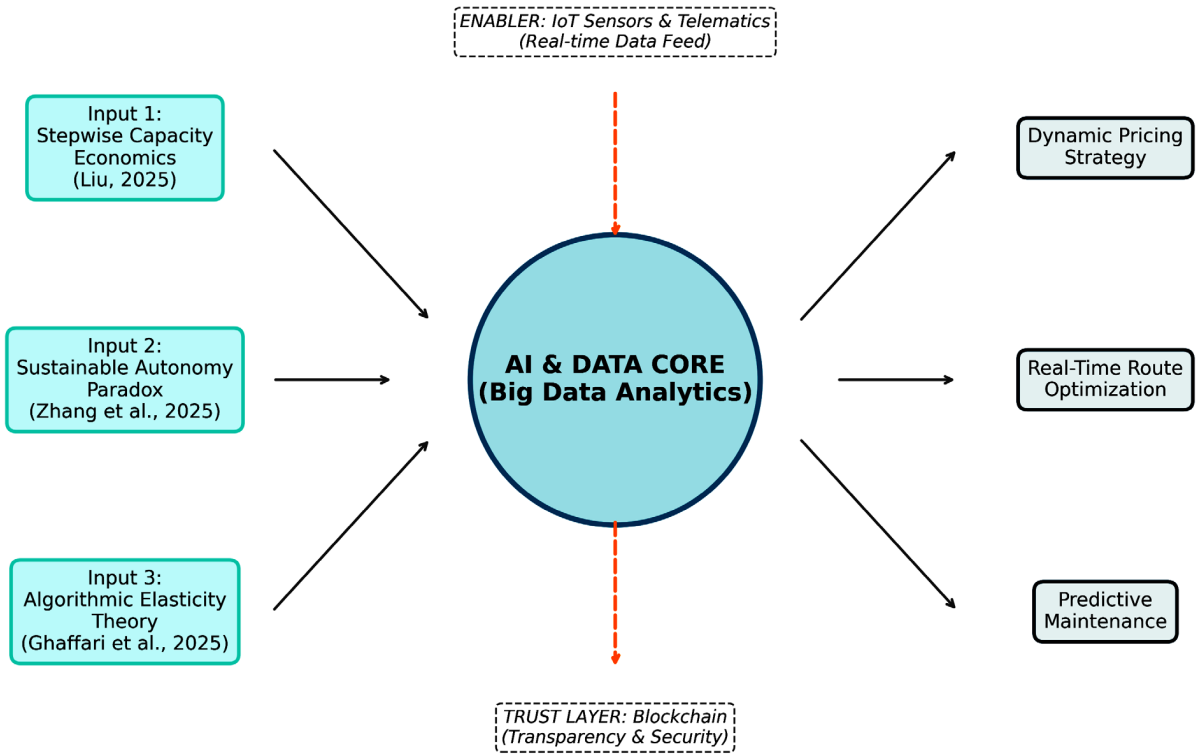

**Figure 2. The Dynamic-Sustainable Cost Planning Framework**

*(Source: Conceptualized by Darwisman, based on Waller et al., 2026; Wang & Liu, 2025; Dutta et al., 2020)*

A schematic diagram showing the convergence of three inputs: (1) Stepwise Cost Structure, (2) Green Autonomy, and (3) Dynamic Revenue. These inputs feed into a central "AI & Data Core" which processes real-time data from IoT sensors to output optimized decisions for Pricing, Routing, and Maintenance.

### 5.4. Managerial Implications

For practitioners, **Darwisman's theory** dictates a radical change in strategic orientation. First, regarding budgeting, organizations must transition from static annual budgeting processes to dynamic, rolling forecasts enabled by Artificial Intelligence, as demonstrated by Almazroi and Ayub (2023), to better adapt to market volatility. Second, in terms of pricing, companies should adopt dynamic pricing models that can protect profit margins against the sudden stepwise cost jumps identified in the new theoretical framework. Third, the asset strategy must shift its focus from merely reducing Operational Expenditures, such as cutting driver wages, to optimizing Capital Expenditures efficiency by investing in Autonomous Electric Vehicles and Artificial Intelligence infrastructure. Finally, concerning maintenance, managers must utilize predictive maintenance tools, as highlighted by Sengupta et al. (2025) and Ujlacka and Konečny (2025), to prevent unplanned downtime, which critically disrupts the synchronization required by the stepwise capacity model.

## 6. CONCLUSION

This study has systematically synthesized high-impact literature from 2018 to 2025, including forward-looking 2026 references, to reconstruct Transportation Cost Planning Theory. We conclude that the era of static, linear cost planning has effectively ended. The new **Dynamic-Sustainable Cost Planning Theory**, formulated by **Samuel Darwisman**, posits three core tenets. First, cost linearity is obsolete; planning must utilize stepwise functions and Mixed-Integer Programming to accurately reflect real-world capacity constraints, as validated by Liu (2025). Second, Artificial Intelligence is the new planner; Machine Learning is no longer optional but essential for dynamic pricing and real-time route optimization, shifting the focus from cost minimization to value maximization, a shift supported by Ghaffari et al. (2025) and Peng et al. (2025). Third, sustainability is profitable; the integration of Autonomous Electric Vehicles and green logistics practices reduces long-term operational costs, making sustainability a key economic driver rather than a mere constraint, as demonstrated by Zhang et al. (2025) and Zhong et al. (2023).

**Novelty Statement:**

Unlike previous theories that treated technology merely as a support tool, **Darwisman's framework** positions Artificial Intelligence, the Internet of Things, and Autonomy as the fundamental architects of the cost structure itself. It provides the first theoretical integration of stepwise costs, algorithmic pricing, and autonomous fleet economics into a single, cohesive framework.

**Future Research Directions:**

While this study establishes a robust theoretical framework, future research should focus on the economic resilience of these AI-driven models against emerging cyber threats, as highlighted by Sengupta et al. (2025). Additionally, empirical studies are needed to quantify the implementation costs of "Fleet Coalitions" and to explore the regulatory frameworks required for autonomous fleet cost-sharing in urban environments. Furthermore, the intersection of Generative AI and logistics planning presents a fertile ground for future theoretical expansion.